\documentclass[reprint,
superscriptaddress,
 amsmath,amssymb,
 aps,
pra,nofootinbib
]{revtex4-1}

\usepackage{graphicx}
\usepackage{dcolumn}
\usepackage{bm}
\usepackage{hyperref}
\hypersetup{colorlinks = true, linkcolor = [rgb]{0.19411,0.51882,0.667058}, urlcolor = [rgb]{0.125490,0.29542,0.1647058}, citecolor = [rgb]{0.75882,0.37411,0.14117}}

\graphicspath{{../images/}}

\usepackage[toc,page]{appendix}

\newcommand{\nn}{\nonumber \\}
\newcommand{\bra}[1]{\langle{#1}|}
\newcommand{\ket}[1]{|{#1}\rangle}
\newcommand{\braket}[2]{\langle{#1}|{#2}\rangle}
\def\togli#1{}
\def\tr{\mbox{Tr}}

\usepackage{braket}

\def\>{\rangle}
\def\<{\langle}

\usepackage[bottom]{footmisc}

\begin{document}

\title{Ancilla-assisted schemes are beneficial for Gaussian state phase estimation}

\author{Zixin Huang}

\affiliation{ Department of Physics \& Astronomy, University of Sheffield, UK }

\author{Chiara Macchiavello}

\affiliation{Dip. Fisica and INFN Sez. Pavia, University of Pavia, via Bassi 6, I-27100 Pavia, Italy}

\author{Lorenzo Maccone}

\affiliation{Dip. Fisica and INFN Sez. Pavia, University of Pavia, via Bassi 6, I-27100 Pavia, Italy}

\author{Pieter Kok}
\affiliation{ Department of Physics \& Astronomy, University of Sheffield, UK }

\begin{abstract}
We study interferometry with Gaussian states and show that an ancilla-assisted scheme outperforms coherent state interferometry for all levels of loss.
We also compare the ancilla-assisted scheme to other interferometric schemes involving squeezing, and show that it is the most advantage in the high-loss, high photon-number regime. In fact, in the presence of high loss, it out-performs many other strategies proposed to date. 
We find the optimal measurement observable for each scheme discussed.
We also find that with the appropriate measurement, the achievable precision of the proposal by Caves [Phys. Rev. D 23, 1693 (1981)] can be improved upon, and is less vulnerable to losses than previously thought.
\end{abstract}

\date{\today}
\maketitle

\section{Introduction}

Quantum metrology describes strategies which allow the estimation precision to surpass the limit of classical approaches \cite{giovannetti2004quantum,PhysRevLett.96.010401,PhysRevLett.98.090501}. When the system is sampled $N$ times, there are different strategies \cite{rafal} which will allow one to achieve the Heisenberg limit, where the variance of the estimated parameter $\Delta^2 \varphi$ scales as $1/N^2$. All of these are equivalent when the systems are noiseless. However, in the presence of noise, these strategies are shown to be inequivalent, where entanglement and the use of ancillae are shown to improve the precision of the estimation \cite{guta}. 

One strategy to reduce the effect of noise is to use an ancillary system that is entangled with the probes but does not participate in the estimation \cite{rafal}.
For qubit systems, it has been shown for many channels that the ancilla is useful for all levels of the noise parameter \cite{PhysRevA.94.012101,kolodynski2013efficient}, but not for bosonic loss channels in the small $N$ limit \cite{PhysRevA.97.032333}.

In optical interferometry, a coherent-light-based strategy is most commonly used but its sensitivity for phase estimation is shot-noise limited, namely 
$\Delta \varphi^2 \geq N^{-1}$. If one needs to achieve a 
finer precision given a finite amount of resources, one has to resort to interferometry with nonclassical states, such as the coherent squeezed state \cite{PhysRevD.23.1693}, two-mode squeezed-vacuum \cite{PhysRevLett.104.103602,PhysRevA.95.053837}, NOON states \cite{Dowling2008}, and squeezed vacuum states \cite{PhysRevA.99.042122,PhysRevA.98.023803,PhysRevA.94.023834}. For works relating to Gaussian state quantum metrology, see, e.g Refs.~\cite{PhysRevA.98.012114,demkowicz2015quantum,vsafranek2015quantum,PhysRevA.85.010101}.

\begin{figure}[b]
\includegraphics[trim = 0cm 0cm 0cm 0cm, clip, width=1.0\linewidth]{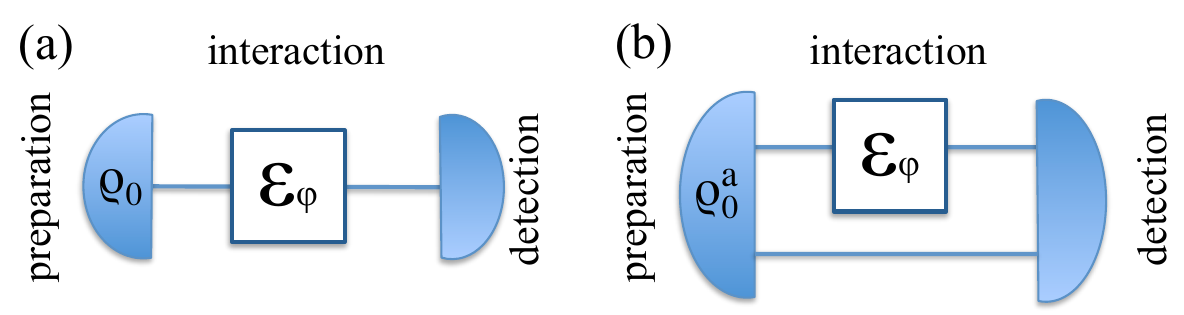}
\caption{\label{fig:anc} (a) Conventional quantum parameter estimation: the state is prepared, it interacts with the probed system through a noisy channel $\mathcal{E}_\varphi$ , followed by the measurement.
(b) Entanglement-assisted parameter estimation: ancillary systems are employed that do not interact with the system.}
\end{figure}

In this paper, we consider a single-parameter estimation task, where the goal is to determine the relative phase shift between two arms of a Mach-Zenhder interferometer.
For the coherent squeezed vacuum \cite{PhysRevD.23.1693}, loss has been considered in Refs.~\cite{PhysRevA.81.033819,PhysRevA.88.041802}, however, the measurement is restricted to photon counting. The SU(1,1) interferometer has been extensively studied (see for example Refs.~\cite{PhysRevLett.119.223604, PhysRevA.86.023844,PhysRevA.95.012109,PhysRevA.98.023803} and here we show that in the high noise regime, at least for some parameter space, the ancilla-assisted scheme out-performs both the  SU(1,1) and the coherent state. Another ancilla-assisted scheme has been considered in Ref.~\cite{PhysRevA.99.043815}. The difference between our work and Ref.~\cite{PhysRevA.99.043815} is that they discard the ancilla at the measurement stage, whereas we allow for the most general measurement that includes the ancilla.

\color{black}

The structure of the paper is as follows. After defining the preliminaries in Sec.~\ref{sec:prelim}, in Sec.~\ref{qmet} we will summarise the key concepts in quantum metrology. We describe the tools we use to calculate the quantum Fisher information (QFI) of arbitrary multimode Gaussian states, as well as the method we use to find the measurement observables.
 We then compare the QFI for the different states we consider, and find the
optimal measurement observable.  The results are presented in Sec.~\ref{results} where we compare the ancilla-assisted state to various others discussed in the literature. Appendix~\ref{append_cal} includes an example calculation for the phase variance.
In Appendix \ref{append_a} we derive the bound for the coherent state. Appendix\ref{sec:width} considers the width of the Fisher information peak with respect to the mean photon number.


\color{black}

\section{Preliminaries}
\label{sec:prelim}
For an $n$-mode bosonic state described by quadrature operators $\hat R=(x_1, x_2,...x_n, p_1,...,p_n)$, $\hat R$ satisfies
\begin{align}
[R_k, R_l] = i \Omega_{k,l} \qquad 
\Omega := \left(
\begin{matrix} 
0 & 1 \\
-1 & 0 
\end{matrix} \right)\otimes \openone
\end{align}
\noindent where $\openone$ is the $n\times n$ identity matrix. For a Gaussian state, given density matrix $\hat \rho$, its properties are completely specified by the
first- and second-moment of the state
\begin{align}
\mu_k = \text{Tr}( R_k \hat \rho), \color{black} \qquad V =  \text{Tr}[\{R_k-\mu_k, R_l-\mu_l \} \hat\rho]
\end{align}
\noindent where $\{A ,B \} = \frac{1}{2}(AB+BA)$ denotes the anticommutator. We have
\begin{eqnarray}                     
         x_i =&& (b_i + b_i ^\dagger)/\sqrt 2 \nn
         p_i =&& -i(b_i - b_i ^\dagger)/\sqrt 2
\end{eqnarray}
\noindent where $b_i ^\dagger, b_i$ are the creation and annihilation operators for mode $i$.

A concept that is useful is the \textit{fidelity} between two states
\begin{align}
\mathcal{F}(\hat \rho_1,\hat\rho_2) = 
\text{Tr}\left[\sqrt{ \sqrt{\hat\rho_1} \hat\rho_2 \sqrt{\hat\rho_1} }\right],
\end{align}
\noindent 
which is shown to be \cite{PhysRevLett.115.260501}
\begin{align}
F(\hat \rho_1, \hat \rho_2) &= \mathcal{F}_0(V_1,V_2) \exp \left[ - \frac{1}{4} \delta_u^T (V_1+V_2)^{-1} \delta   \right]. \label{eq:fid1}\\
  \mathcal{F}_0 (V_1,V_2)   &=  \frac{\mathcal{F}_\text{tot}}{\text{det}(V_1+V_2)^{1/4}}, \\
  \mathcal{F}^4_\text{tot}   &= \det \left[ 2\left( \openone + \sqrt{\frac{V_\text{aux} \Omega}{4}} +  \openone \right) V_\text{aux}\right],   \\
  V_\text{aux} &= \Omega^T(V_1 + V_2)^{-1} \left( \frac{\Omega}{4} + V_2 \Omega V_1  \right),
  \label{eq:fid}
\end{align}
with $\delta = \mu_1 -\mu_2$.

The states considered in this paper include the coherent state, the one- and two-mode squeezed vacuum.
A coherent state with mean photon number $|\alpha|^2$ can be written in the Fock basis,
\begin{align}
e^{\alpha  a^\dagger - \alpha^* a} \ket{0} = &\ket{\alpha} = e^{-|\alpha|^2/2}\sum_{n = 0}^\infty \frac{\alpha^n}{n!}\ket{n}, \nn 
 \text{Tr}[ a\ket{\alpha}\bra{\alpha}] = & ~\alpha.
\end{align}
 \noindent Without loss of generality, we will consider $\alpha \in \mathbb{R}$ in the rest of the paper.
 The one-mode squeezed vacuum with squeezing parameter $r$ can be written as
\begin{align}
  &e^{r/2(a^{\dagger2} e^{i\theta}- a^{2} e^{-i\theta})} \ket{0}  
  \end{align}
 \noindent and with $\theta=0$, it can be written in the Fock basis 
  \begin{align}
&\frac{1}{\sqrt{\cosh r}}\sum_{n=0}^\infty \left( \frac{\tanh r}{2} \right)^n \frac{\sqrt{(2n)!}}{n!}\ket{2n},
\end{align}
\noindent and has mean photon number $\sinh^2 r$.
Similarly, choosing a real squeezing parameter, the two-mode squeezed vacuum has an analogous representation:
\begin{align}
  &e^{r(a_1^{\dagger} a_2^\dagger - a_1 a_2 )} \ket{0,0}  \nn
= &\frac{1}{{\cosh r}}\sum_{n=0}^\infty (\tanh r)^n \ket{n,n }.
\end{align}
\noindent The mean total photon number of a TMSV is $2\sinh^2 r$, and the energy is equally shared
between the two modes. \color{black}

\section{Quantum metrology}
\label{qmet}

A quantum parameter estimation process is composed of three stages, see Fig.~\ref{fig:anc}(a):  
\begin{enumerate}
\item The probe system is initialized in the preparation.
\item  The probes interact with the system to be sampled, where the interaction encodes the parameter on the probes.
\item  The measurement stage, the probes are measured and the outcome is processed to yield the parameter estimate.
\end{enumerate}

\noindent Entanglement-assisted quantum metrology (depicted in Fig~\ref{fig:anc} (b)) refers to the scenario where the probes are entangled with an ancilla that does not participate in the sampling stage. Then at the measurement stage a joint measurement is performed on probes and the ancilla. 
\color{black}

Here our parameter of interest is the change of optical path length of the mode $k$, with the unitary 
\mbox{$U_{\varphi}^k= e^{i \varphi a_k^\dagger a_k}$}, and we detail the increase in achievable precision in the presence of an entangled ancilla. The goal is to determine the parameter $\varphi$ by performing the best possible POVM measurement on $\hat \rho_\varphi$.  For the bosonic loss channels that we consider, $U_\varphi$ commutes with the action $\mathcal{E}[\cdot]$ of the noise:
\begin{align}
\hat \rho_\varphi = U_\varphi \mathcal{E}[\hat \rho_0] U_\varphi^\dagger = \mathcal{E}[U_\varphi\hat \rho_0U_\varphi^\dagger] ,
\end{align}
\noindent note that we have dropped the subscript $k$ for brevity.
 Not only can we
exchange the unitary and the noise in this case, but the noise and the
phase shift might also act simultaneously.

In conventional quantum parameter estimation, the initial state is $\hat \rho_0$ and it passes through a quantum channel $\mathcal{E}_\varphi$ which encodes the parameter $\varphi$,
\begin{align}
\hat \rho_0 \rightarrow \mathcal{E}_\varphi(\hat \rho_0) =\hat\rho_\varphi.
\end{align}

\noindent 
Here we allow for the addition of an ancillary state, which may be entangled with the initial state, but it does not participate in the interaction, nor does it experience the noisy channel
\begin{align}
\hat \rho_{0}^a \rightarrow \left(\mathcal{E}_\varphi \otimes \openone \right)[\hat \rho_{0}^a] = \hat \rho_{\varphi}^a.
\end{align}

For the remainder of the paper, we will compare schemes with the same photon number entering the noisy quantum channel that cross the phase shift $U_\varphi$, and study the precision as a function of the loss parameter.

The ultimate precision of the estimation is given by the quantum Cramer-Rao (QCR) bound \cite{holevo,helstrom,caves,caves1}. It is a lower bound to the
the variance of the estimation of a parameter $\varphi$ encoded onto a state $\hat \rho_\varphi$ by an interaction $\mathcal{E}_\varphi$.
For unbiased estimators, $\Delta \varphi ^2 \geqslant 1/\nu
  J(\hat \rho_\varphi)$, where $\nu$ is the number of times the estimation
is repeated, and $J(\hat \rho_\varphi)$ is the quantum Fisher information (QFI)
associated with the global state $\hat \rho_\varphi$ of probes and ancillae
(after the interaction $\mathcal{E}_\varphi$ with the probed system). When there is a unique most probable estimate, the bound is achievable in the asymptotic limit that $\nu \rightarrow \infty$.

The definition for QFI we use here is based on the distinguishability of the states:
%
\begin{eqnarray}
J(\hat \rho_\varphi)= \frac{8[1-\mathcal{F}(\hat \rho_\varphi, \hat \rho_{\varphi+d\varphi})]}{d\varphi^2}.
\end{eqnarray}
%
We calculate $\mathcal{F}(\hat \rho_\varphi, \hat \rho_{\varphi+d\varphi})$ using Eqs.~\eqref{eq:fid1}-\eqref{eq:fid}, and evaluate the QFI numerically where the analytical expression becomes intractable.

For pure states and unitary processes $U_\varphi = e^{i \hat G \phi}$, the QFI is equal to 4 times the variance of the generator \cite{paris2009quantum},
\begin{align}
J(\varphi) = 4 \left(\braket{\hat G^2} - \braket{\hat G}^2 \right).
\end{align}


We will now move on to discussing the measurement.
%
%
The QFI provides us with an upper bound to the achievable precision, but does not indicate the optimal measurement. 

For a given measurement $M$ whose outcomes are $\{m_i\}$ occuring with probability $\{p_i\}$, its Fisher information
\begin{align}
I(\varphi) =\sum_i p_i \left(\frac{\partial }{\partial \varphi}\log[p_i] \right)^2,
\end{align}
\noindent and the variance of $M$ is upper bounded by $I(\varphi)^{-1}$.
The measurement is optimal if the Fisher information is equal to the QFI.

In the most general case, for single-parameter estimation, the optimal {estimator} is formally given by \cite{paris2009quantum}
\begin{eqnarray}
\hat M_{\text{opt}} = \varphi + {\hat L }{J^{-1}(\rho_\varphi)}.
\label{eq:optimal_estimator}
\end{eqnarray}
\noindent where $\hat L$ is the symmetric logarithmic derivative (SLD), defined as
\begin{eqnarray}
\frac{\partial \rho}{\partial \varphi} = \frac{1}{2}(\hat L \hat \rho_\varphi + \hat\rho_\varphi \hat L).
\end{eqnarray}
\noindent We can see that Eq.~\eqref{eq:optimal_estimator} is true from the error-propagation formula
\begin{align}
\text{Tr}[\hat M_{\text{opt}} \hat \rho ] = \varphi, \qquad
\text{Tr}[\hat M^2_{\text{opt}} \hat \rho ] = \varphi^2 + \frac{\text{Tr}[\hat \rho \hat L^2]}{J^2(\varphi)},
\end{align}
\noindent and thus $\braket{\Delta\hat  M^2_\text{opt} } = J(\varphi)^{-1}$.

\color{black}

 Therefore finding the SLD gives insights into the measurement that one needs to perform. Since this observable depends on the parameter $\varphi$ to be estimated, the experimenter must use an estimate $\theta$ in place of $\varphi$ and use a feedback strategy \cite{PhysRevLett.85.5098,PhysRevLett.104.063603,PhysRevLett.107.233601,PhysRevA.95.053837} to adjust $\theta \rightarrow \varphi$.

Finding $\hat L$ analytically for an arbitrary mixed multimode Gaussian state is non-trivial. For pure states,
following the result by Monras, $\hat L$ is given by \cite{monras2013phase} 
\begin{eqnarray}
\hat L_\varphi =\sum_{i,j}&& -\frac{1}{2} \partial_\varphi(\Gamma^{-1})_{i,j}   \{R^i-\mu^i,R^j-\mu^j\} \nn
                      && + 2 (\partial_\varphi \mu^i) (\Gamma^{-1})_{i,j}(R^j-\mu^j) \nn
                     &&+\text{constant}
           \label{eq:sld}
\end{eqnarray}
Here $ \Gamma = 2\tr[\{R^j-\mu^j,R^k-\mu^k\}\rho]$.
We use Eq.~\eqref{eq:sld} as the basis to derive the measurement observables, but we omit the constant term.

\section{Results}\label{results}

The main results of this paper are: in Sec.~\ref{anc_to_coh}
 we show that an ancilla-assisted scheme can beat the coherent state (which is considered robust against losses 
 \footnote{The coherent state is robust against losses in the sense that a pure loss channel does not change the quadrature variance of the state, i.e. the state remains pure after interaction with the channel, only the amplitude is reduced.}) 
 for all levels of the loss parameter. 
 Sec.~\ref{coherent_squeezed} shows that, if we consider the optimal measurement as opposed to photon counting, the precision of the coherent squeezed vacuum proposed in 
 Ref.~\cite{PhysRevD.23.1693} can be improved. 
 In Sec~\ref{all_schemes} we compare the ancilla-assisted schemes to 
 some of the most popular schemes including the squeezed vacuum, coherent squeezed vacuum etc, and the SU(1,1) interferometer
  \cite{PhysRevA.86.023844,PhysRevLett.119.223604}. We show that they can be improved with a better measurement, and that the ancilla-assisted strategy 
can still out-perform the states in (2) in the high loss regime, at least for part of the parameter space.

One of the benchmarks we compare against is the bound for the coherent state. It is derived as follows. 
 A coherent state $\ket{\alpha}$ (with mean photon number $\alpha^2$) passes 
through the first 50:50 beam splitter of the Mach-Zenhder interferometer (MZI). It then goes through a lossy channel with transmitivity $\eta$ and experiences the phase shift $U_\varphi$. \color{black} The QFI for the state is 
\begin{align}
J_\text{coh}(\varphi) &= 4 \left( \braket{ \hat G^2}-\braket{ \hat G}^2 \right), \qquad \hat G = a^\dagger a \nn
                      &=  2 \eta \alpha^2.
\end{align}
\noindent Note that the mode which contains the phase shift has an average photon number $\alpha^2/2$. We show how this bound can be achieved in Appendix~\ref{append_a}.

\color{black}

\subsection{ Ancilla-assisted scheme compared to the coherent state}
\label{anc_to_coh}

\begin{figure}[tbh]
\includegraphics[trim = 0cm 0cm 0cm 0cm, clip, width=0.8\linewidth]{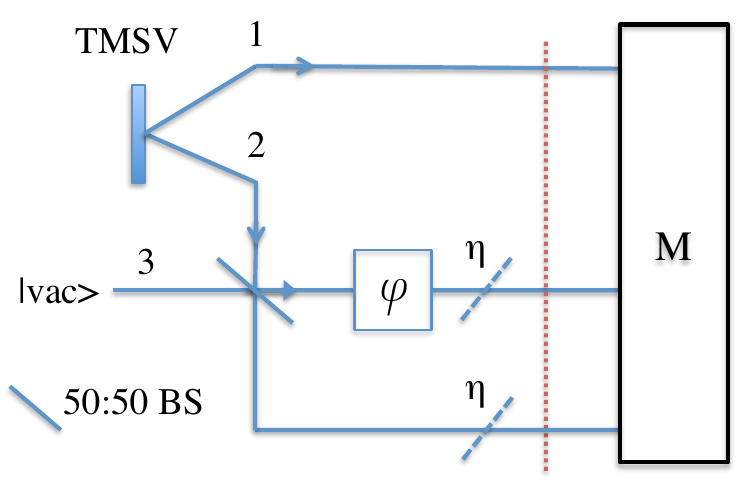}
\includegraphics[trim = 0cm 0cm 0cm 0cm, clip, width=0.9\linewidth]{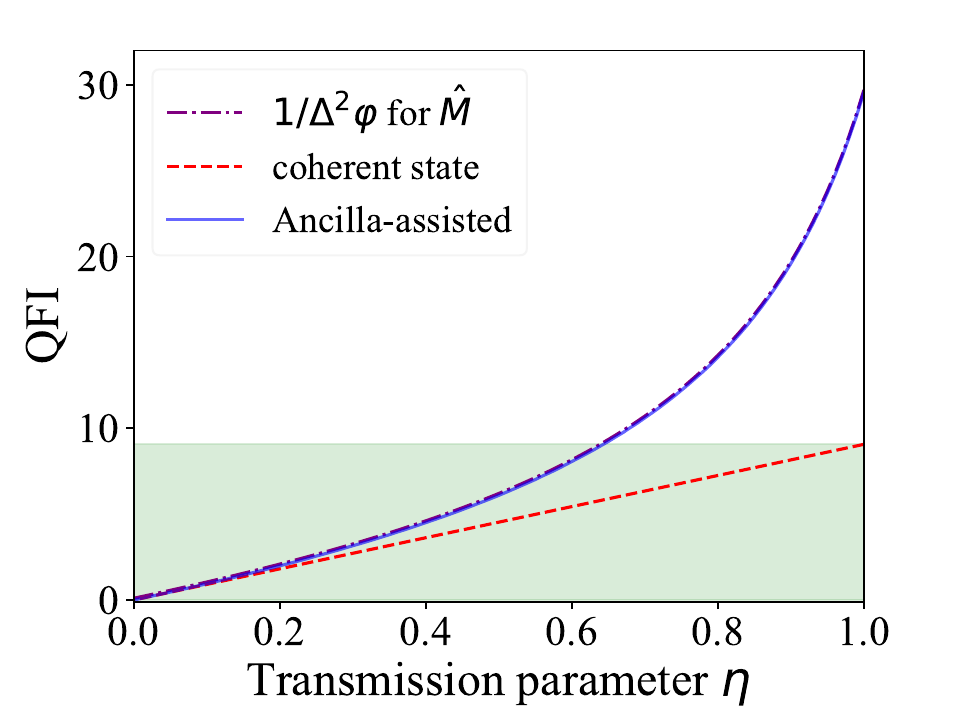}
\caption{\label{fig:tmsv} (Top) Ancilla-assisted scheme: modes 1 and 2 are a two-mode squeezed vacuum state where mode 1 is the ancilla, and mode 2 is input into a Mach-Zenhder interferometer. Modes 2 and 3 are lossy.
(Bottom) QFI for the TMSV with squeezing parameter $r = 1.5$ (blue solid line), QFI of the coherent state with the same mean photon number (dashed red line), bound for the coherent state (green solid line), and $1/\Delta^2\varphi$ for the measurement $\hat M_{\text{anc}}$, at $\varphi = 0$ (purple dotted dashed line). For all level of the loss parameter, the ancilla-assisted scheme is advantagous. The top of the shaded region denotes the coherent state bound, and is for guide of the eye.}
\end{figure}

The first scheme we consider is depicted in Fig.~\ref{fig:tmsv} (top). The initial state is a two-mode squeezed vacuum (TMSV) acting upon by a 50:50 beam splitter, 
\begin{eqnarray}
\text{BS}_{23}\exp \left (r a_1 a_2 - r a_1 ^\dagger a_2 ^\dagger \right) \ket{0,0,0}_{1,2,3}
\end{eqnarray}
where $r$ is the squeezing parameter, and the subscripts in BS$_{23}$ denotes that the beam splitter acts on modes 2 and 3. Here we take $r$ to be real.

Mode 2 is fed into the input of a 
Mach-Zenhder interferometer (MZI), where the third mode experiences a relative phase shift 
$a_3^\dagger \rightarrow a_3^\dagger e^{i \varphi} $. The MZI is assumed to have losses on both arms, with
transmission parameter $\eta$. The ancilla in mode 1 is assumed to be noiseless.
%
The total mean photon number of a TMSV is $2 \sinh^2 r$, therefore the mean photon number going through the phase shift, without loss, is $\frac{1}{2} \sinh^2 r$.

For the noiseless case ($\eta=1$), using Eq.~\eqref{eq:sld}, the SLD is 
\begin{align}
\hat M_{\text{pure}} &=i  \cosh r \sinh r \left (b_1 b_3 - b_1^\dagger b_3^\dagger \right)/\sqrt2 \nn
       &+ i   \sinh^2 r \left(b_3b_2^\dagger - b_2 b_3^\dagger \right)/2.
\end{align}

The QFI for mixed state is sometimes calculated numerically, given the difficulty of obtaining a
general analytic expression for it. 
We plot the QFI for the two-mode squeezed vacuum with $r=1.5$  (solid blue line) and the coherent state with $\alpha = \sinh 1.5$ (dashed red line). We see that the ancilla-assisted scheme out-performs the coherent state.

We found that measuring the observable $\hat M_{\text{anc}}$ can beat the shot noise limit, 
\begin{eqnarray}
\hat M_{\text{anc}} &=&i \sqrt \eta \cosh r \sinh r \left(b_1 b_3 - b_1^\dagger b_3^\dagger \right)/\sqrt2 \nn
       &+& i  \eta \sinh^2 r \left(b_3b_2^\dagger - b_2 b_3^\dagger \right)/2.
\label{eq:anc_obs}
\end{eqnarray}

\noindent 
The scaling factors $\sqrt \eta$ and $\eta$ in the first and second term respectively are not present in the pure state case from Eq.~\eqref{eq:sld}. They were added to weigh the terms, which gives a smaller variance.
\color{black} The expectation value of $\hat M_{\text{anc}}$ is
\begin{eqnarray}
 \braket{\hat M_\text{anc}}&=&\frac{1}{4} \eta  \sinh ^2(r) \sin (\varphi ) 
 \left[-\eta +(\eta -2) \cosh (2 r)-2 \right]. \nn 
\end{eqnarray}

Using error propagation:
\begin{eqnarray}
\Delta^2\varphi &=& \frac{{\Delta^2M_{\text{anc}}}}{ (\partial \braket{M_{\text{anc}}}/\partial \varphi)^2} \nn
 &=&\frac{x}{8 \eta  \sinh ^2(r) \cos ^2(\theta ) (-\eta +(\eta -2) \cosh (2 r)-2)} \nn
x&=& -3 \eta ^2 \cos (2 \varphi )+3 \eta ^2-2 \eta  \cos (2 \varphi )-6 \eta \nn
 & & +4 \cosh (2 r) (\eta ^2 \cos (2 \varphi )-\eta ^2 \nn
 & & +2 \eta -2)+2 (\eta -2) \eta  \cosh (4 r) \sin ^2(\varphi )-8.
 \label{eq:varphitmsv}
\end{eqnarray}

\noindent The minimum phase variance is achieved at $\varphi=0$,
\begin{eqnarray}
\Delta^2\varphi_\text{min}=\frac{ (\eta +(1-\eta) \cosh (2 r)+1)}{\eta  \text{sinh}^2(r)(\eta +(2-\eta) \cosh (2 r)+2)}. \nn
\label{eq:tmsv}
\end{eqnarray}

The fact that the precision depends on the estimated parameter is typical in quantum metrology and one can use feedback strategies to find the optimal working point \cite{PhysRevLett.85.5098,PhysRevLett.104.063603,PhysRevLett.107.233601,PhysRevA.95.053837}.

Comparing the quantity in Eq.~\eqref{eq:tmsv} to the bound for coherent states with the same mean photon number (which is $2 \eta \alpha^2$), it
is always smaller than $2 \eta \sinh^2(r)$. \color{black}
For $\eta=1$, it is easy to see that $\hat M_\text{anc}$ is optimal, achieving the QCRB 
\begin{align}
\Delta^2\varphi = 1/[\sinh^2(r) (\sinh^2 (r)  +2)].
\end{align}

In Fig.~\ref{fig:tmsv}, we see that the inverse of Eq.~\eqref{eq:tmsv} overlaps with the QFI, which we numerically verified for a range of values of $r$. 
This strongly suggests that $\hat M_\text{anc}$ is optimal, suggesting that the inverse of Eq.~\eqref{eq:tmsv} is equal to the QFI. 
\vspace{5mm}

\subsection{The coherent + squeezed state scheme with optimal measurement}
\label{coherent_squeezed}

\begin{figure}[tbh]
\includegraphics[trim = 0cm 0cm 0cm 0cm, clip, width=0.8\linewidth]{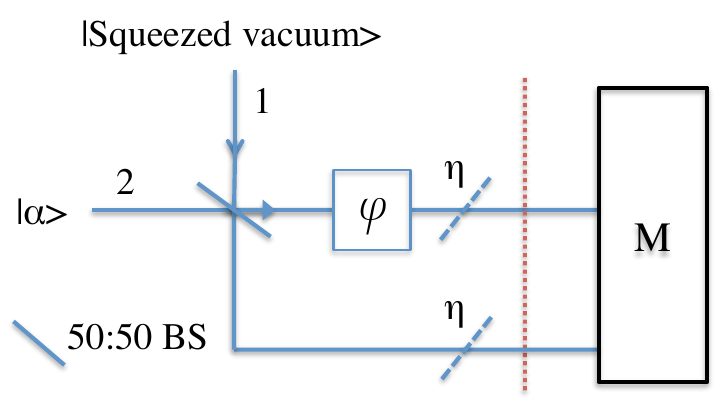}
\includegraphics[trim = 0cm 0.2cm 0cm 0cm, clip, width=1.0\linewidth]{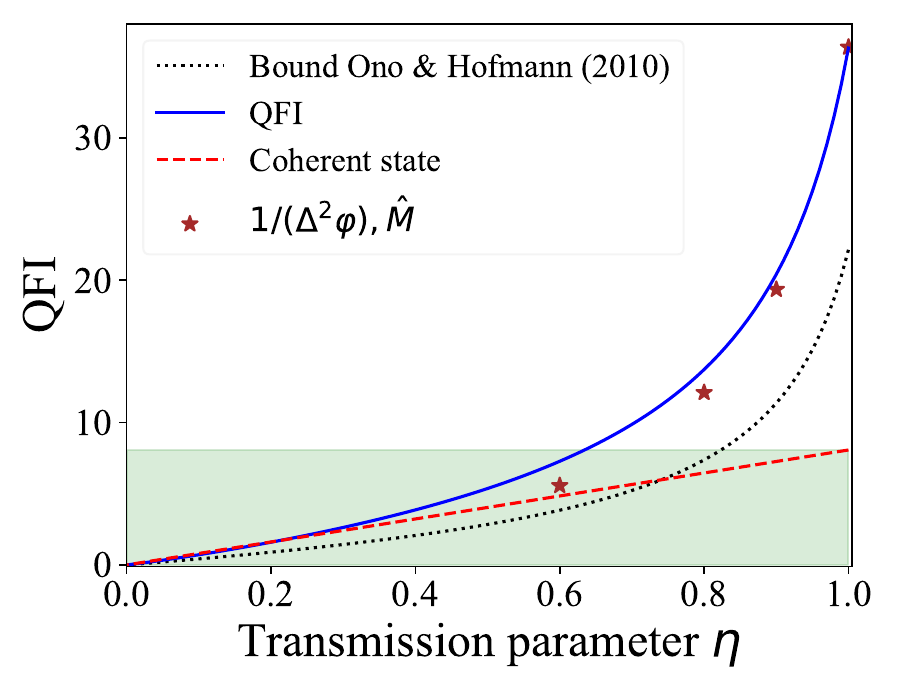}
\caption{\label{fig:caves_qfi} (Top) Scheme proposed by Caves, where one input port is the coherent state and the other is the squeezed vacuum. (Bottom) QFI for the scheme with parameters $r=1.15, \alpha^2= \sinh^2r$. In this plot we show: QFI of the scheme (blue solid line), and $1/\Delta^2\varphi$ for the state using $\hat M$ at $\varphi=0$ (purple stars), the Fisher information for photon counting (black dotted line) from Ref.~\cite{PhysRevA.81.033819}, the Fisher information using $\hat M$ at $\varphi = 0$ (purple stars) and the coherent state (red dashed line). The top of the shaded region denotes the coherent state bound, and is for guide of the eye.}
\end{figure}

Next, we consider the scheme proposed by Caves (Fig.~\ref{fig:caves_qfi}) where one input to the MZI is a coherent state $\ket{\alpha}$ and the other is the squeezed vacuum.
%
The scheme in the presence of loss has been considered in Refs~\cite{PhysRevA.81.033819,PhysRevA.88.041802}, both of which calculate the achievable precision using estimators
that are based on measuring photon-number-differences between the two modes, i.e. they base their calculations on the Jordan-Schwinger formalism \cite{demkowicz2015quantum}. 
\color{black}

The input state is
\begin{eqnarray}
\text{BS}_{12}\exp \left( r \frac{a_1^2}{2} - r \frac{a_1^\dagger}{2} \right)\ket{0,\alpha}_{1,2}.
\end{eqnarray}
\noindent  The mean photon number of a squeezed vacuum is $\sinh^2 r$, and the mean photon passing through the phase shift is $\frac{1}{2}(\alpha^2 + \sinh^2 r).$ 
\color{black}
\noindent The Fisher information for measuring the photon number is known to be \cite{PhysRevA.81.033819,PhysRevA.88.041802,PhysRevLett.100.073601}
\begin{eqnarray}
(\Delta^{2} \varphi)^{-1} &=& \left[\alpha^2e^{2r}+\sinh^2 (r) \right].
\label{eq:cavesold}
\end{eqnarray}

\noindent While this is the precision achievable with that measurement, this is not the ultimate achievable precision, i.e. based on the quantum Fisher information. If we calculate the QFI for the lossless case, $4\Delta^2 G$, this gives
\begin{align}
J(\varphi) = &~ 4\left[\braket{(a^\dagger_2 a_2 )^2 } - \braket{(a^\dagger_2 a_2 )}^2\right] \nn
=& ~\alpha ^2+\alpha ^2 e^{2 r}+ \frac{\cosh(4r)}{4} + \frac{\cosh(2r)}{2} -\frac{3}{4}.
\label{eq:qficaves}
\end{align}

\noindent
In the optimal case where we restrict the total photon number in the interferometer, $\alpha$ and $r$ are related by $\sinh^2r = \alpha^2$. Here the value in eq.~\eqref{eq:qficaves} outperforms Eq.~\eqref{eq:cavesold} by more than a factor of $\frac{3}{2}$.  We plot the QFI in Fig.~\ref{fig:caves_qfi}, where the coherent squeezed state has squeezing parameter $r=1.15$.\footnote{Note that the parameters were chosen realistically, but otherwise arbitrarily for graphical convenience, and such that for a given fixed total mean photon number, the highest QFI is achieved.}
 The QFI (solid blue line) is significantly larger than when only photon counting is used (black dotted line, calculated by Ono and Hofmann in Ref.~\cite{PhysRevA.81.033819}).
In this plot, the coherent state used for comparison has a mean photon number of $2\alpha^2$ (red dashed line).

As can be seen in Fig.~\ref{fig:caves_qfi}, the Caves scheme with the optimal measurement out-performs the coherent state up until $\eta \approx 0.6$. However, it does not beat the coherent state at all loss parameters (depending on the squeezing and displacement parameter, which is not obvious from the plot here).

Once again using Eq.~\eqref{eq:sld}, the optimal measurement operator for the lossless case ($\eta =1$) is
\begin{eqnarray}
\hat M=&&\frac{i}{4}( e^{-2 r-2 i \theta }- e^{2 r-2 i \theta })b_2^{\dagger2} + \frac{i}{4}(e^{2 r+2 i \theta }-e^{-2 r+2 i \theta })b_2^2 \nn
&&+\frac{i \alpha   b_1}{2 \sqrt{2}}-\frac{i \alpha   b_1 e^{2 r}}{2 \sqrt{2}}-\frac{i \alpha   b_1^\dagger }{2 \sqrt{2}}+\frac{i \alpha   b_1^\dagger  e^{2 r}}{2 \sqrt{2}} \nn
&&+\frac{3 i \alpha   b_2 e^{i \theta }}{2 \sqrt{2}}+\frac{i \alpha   b_2 e^{2 r+i \theta }}{2 \sqrt{2}}-\frac{3 i \alpha   b_2^\dagger  e^{-i \theta }}{2 \sqrt{2}}-\frac{i \alpha   b_2^\dagger  e^{2 r-i \theta }}{2 \sqrt{2}} \nn
&&-\frac{1}{4} i e^{2 r-i \theta }  b_2^\dagger    b_1-\frac{1}{4} i e^{-2 r-i \theta }  b_2^\dagger    b_1+\frac{1}{2} i e^{-i \theta }  b_2^\dagger    b_1 \nn
&&+\frac{1}{4} i e^{-2 r+i \theta }  b_1^\dagger    b_2+\frac{1}{4} i e^{2 r+i \theta }  b_1^\dagger    b_2-\frac{1}{2} i e^{i \theta }  b_1^\dagger    b_2\nn
&& +\frac{1}{4} i e^{2 r-i \theta }  b_1^\dagger    b_2^\dagger -\frac{1}{4} i e^{-2 r-i \theta }  b_1^\dagger    b_2^\dagger  \nn
&& +\frac{1}{4} i e^{-2 r+i \theta }  b_1   b_2-\frac{1}{4} i e^{2 r+i \theta }  b_1   b_2 %
\label{eq:caves_obs}
\end{eqnarray}
\noindent where $\theta$ is the input parameter to the apparatus which ideally should coincide with, or at least be very close to $\varphi$.
The expectation value of the measurement is
\begin{eqnarray}
\braket{\hat M}=&&\frac{1}{8} \sqrt{\eta } e^{-4 r} \sin (\varphi -\theta ) \nn
           &&\{\sqrt{\eta } \left(e^{4 r}-1\right) \left(4 \alpha ^2 e^{2 r}+e^{4 r}-1\right) \cos (\varphi -\theta ) \nn
           &&+2 e^{2 r} [\left(2 \alpha ^2+1\right) \sqrt{\eta }+e^{4 r} \left(2 \alpha ^2+\sqrt{\eta }\right) \nn
           &&-2 e^{2 r} \left(\alpha ^2 \left(\sqrt{\eta }-3\right)+\sqrt{\eta }\right) ]\}.
\end{eqnarray}
The variance of the measurement is given by a large and un-illuminating expression, which we leave in Appendix~\ref{append_a}. 
The quantum Cramer-Rao bound is achieved at $\varphi-\theta=0$, i.e,
\begin{eqnarray}
\Delta^2\varphi = \left(  \alpha ^2+\alpha ^2 e^{2 r}+ \frac{\cosh(4r)}{4} + \frac{\cosh(2r)}{2} -\frac{3}{4} \right)^{-1}. \nn
\label{eq:caves_qcr}
\end{eqnarray}

Calculating the variance of $\hat M$ for $\eta$ analytically is difficult, but can be done for fixed values. For \mbox{$\eta = \left\{0.6,0.8,0.9,1.0\right\}$}, the minimum variance occur at \mbox{$\varphi-\theta = 0$}, and we plot $1/(\Delta^2\varphi)$ as purple stars in Fig.~\ref{fig:caves_qfi}. As we can see, it performs significantly better than photon counting and almost achieves the QFI (even though this is the optimal measurement only for $\eta$=1).

In general, the Fisher information depends on the (unknown) parameter $\varphi$ to be estimated. The Fisher information as a function of $\varphi$ will typically have a peak where it is large around an optimal estimation point. In Appendix~\ref{sec:width} we examine how sensitive is the Caves-variant scheme's Fisher information to the parameter.
\color{black}

\begin{figure}[tbh]
\includegraphics[trim = 0cm 0cm 0cm 0cm, clip, width=0.75\linewidth]{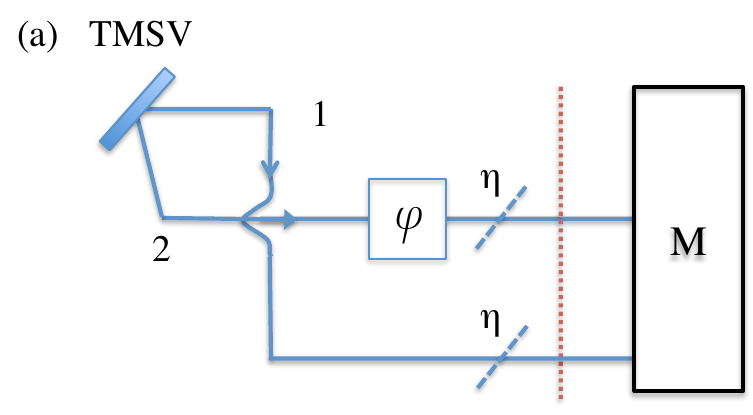}
\includegraphics[trim = 0cm 0cm 0cm 0cm, clip, width=0.75\linewidth]{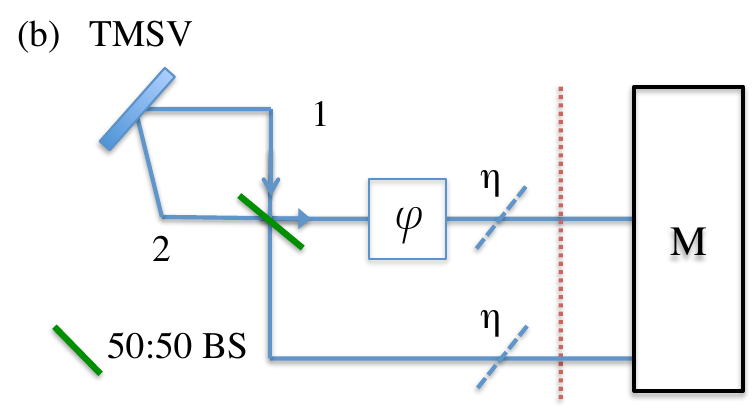}
\caption{\label{fig:su11}: Two versions of the SU(1,1) interferometer: (a) the TMSV is fed straight into the two modes of the interferometer with additional beam splitters, and (b) with a 50:50 beam splitter (BS) drawn in green.}
\end{figure}

We also note that, if one replaces the vacuum in Fig.~\ref{fig:tmsv} with a coherent state $\ket{\alpha}$, this performs less-well than the Caves scheme. For the pure state, its QFI is
\begin{eqnarray}
J_c(\varphi)= \frac{1}{2} \left[4 \alpha ^2 \cosh ^2(r)+\sinh ^2(r) (\cosh (2 r)+3)\right].\nn
\label{eq:hybrid}
\end{eqnarray}
Eq.~\eqref{eq:hybrid} is lower than Eq.~\eqref{eq:qficaves}, although they both have quadratic scaling with the mean photon number.

\subsection{Ancilla-assisted scheme compared with SU(1,1) interferometers}
\label{all_schemes}

We now compare the ancilla-assisted scheme to both the versions of the SU(1,1) interferometer \cite{gao2014bounds} (Fig.~\ref{fig:su11}): (a) the output of a TMSV is fed into
an interferometer, and (b) there is a 50:50 beam splitter at the input \cite{PhysRevLett.104.103602,PhysRevA.95.053837}.  
The state in scheme (b) is transformed into a product state of two one-mode squeezed vacuum. 
Scheme (b) has QFI \cite{gao2014bounds}
\begin{eqnarray}
J(\rho_{\text{omsv}})&=&\frac{4 e^2}{1+d^2-e^2} \nn
 d &=& \eta \cosh(2r)+(1-\eta)\nn
  e&=& -\eta \sinh(2r),
  \label{eq:qfi_su11}
\end{eqnarray}
\noindent which is a factor of 2 larger than the one without the BS, as in scheme (a). This suggests that the disentangled state outperforms the entangled state. In Ref.~\cite{gao2014bounds}, the authors point out that this is due to the fact that the BS turns the TMSV back into two one-mode squeezed vacuum states in their respective mode, and an external phase reference is necessary to extract $\varphi$.  That is, the TMSV is self-sufficient, whereas an extra mode is needed for the single-mode squeezed vacuum; here one may choose to associate the energy cost with the extra mode if necessary.

As derived in Ref~\cite{gao2014bounds}, for the SU(1,1) in scheme (a), the optimal measurement is
$\hat M_{12}=i \left (b_1^\dagger b_2^\dagger - b_1 b_2 \right)$, whilst for scheme (b) it is 
$\hat M_1 = i \left (b_2^{\dagger2} - b_2^2 \right)$.

\begin{figure}[tbh]
\includegraphics[trim = 0cm 0cm 0.5cm 0cm, clip, width=1.0\linewidth]{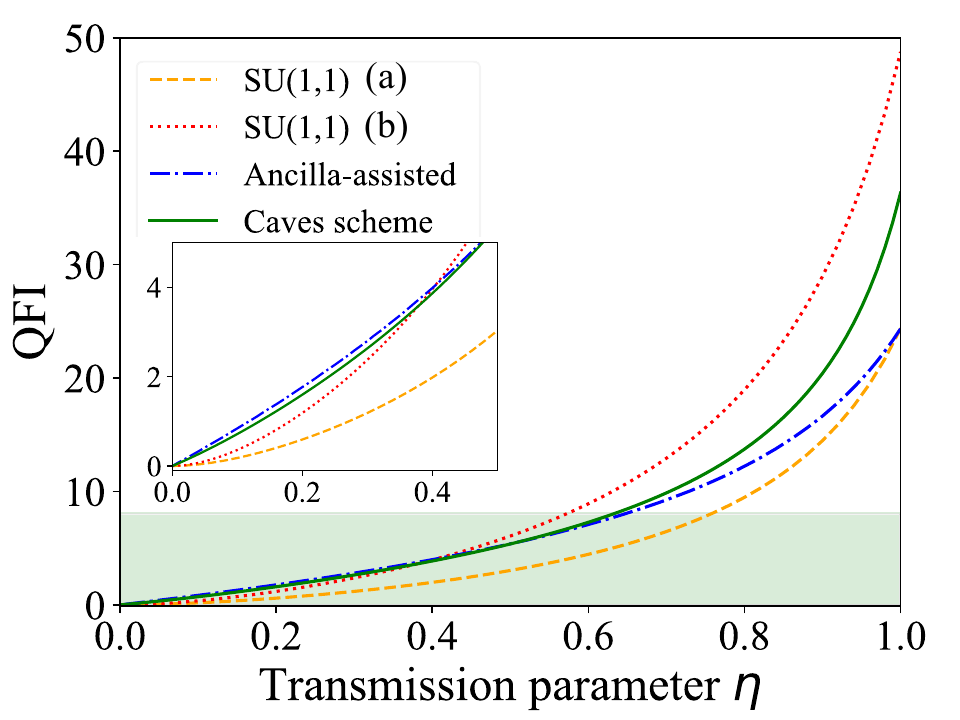}
\caption{\label{fig:all} QFI for an interferometer with mean photon number going through the phase shift $\bar n =\sinh^2r$, $r=1.15$: the
SU(1,1) interferometer with added BS (red dotted line), the SU(1,1) interferometer (yellow dashed line), the Caves scheme (green solid line). The ancilla-assisted state as in Fig.~\ref{fig:tmsv} (blue dotted-doshed line, the squeezing parameter here is $\sinh^{-1}[\sqrt{2 \bar n}]$). The top of the shaded region denotes the coherent state bound, and is for guide of the eye.}
\end{figure}

As seen from Fig.~\ref{fig:all}, in the high-loss regime, the ancilla-assisted strategy out-performs all the other schemes, closely followed by the Caves-variant scheme.  The quantity that is considered fixed during the comparison is the number of photons going through the phase shift. 
%
To clarify, the parameters and total photon number of the states are:
\begin{align}
r = 1.15 , ~ & \alpha = \sinh(1.15) \nn
\text{Caves scheme}: &  ~\sinh^2r + \alpha^2 \nn
\text{SU(1,1) (a) and (b)} :  & ~ 2 \sinh^2 \nn
\text{Ancilla-assisted} : &~ \sinh^2{\left(\sqrt{2} \sinh r \right)}. \nonumber
\end{align}
\noindent Note that the ancilla-assisted scheme has a higher squeezing parameter to account for the fact that one mode of the state is not entering the interferometer.

\color{black}

\begin{figure}
\includegraphics[trim = 1.5cm 0cm 0cm 0cm, clip, width=0.9\linewidth]{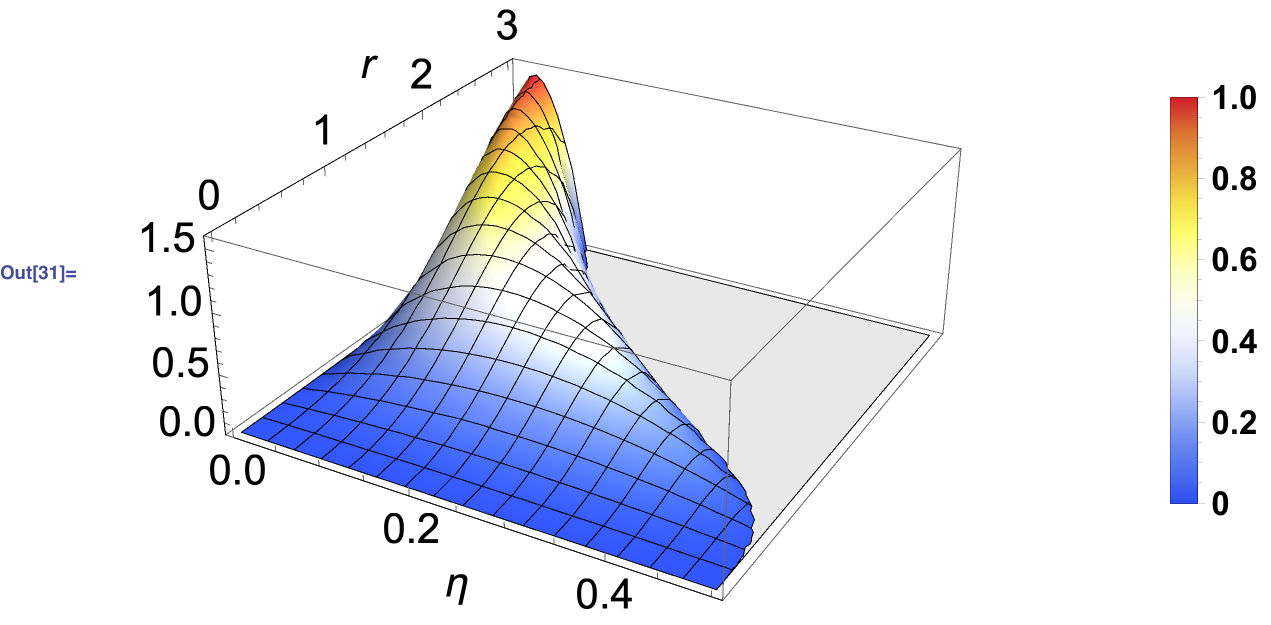}
\caption{\label{fig:su11_anc} The minimum QFI difference between the ancilla-assisted scheme and the SU(1,1) interferometer
in Fig~\ref{fig:su11} (b) for a range of values of the squeezing parameter. For clarity we show the poitive region only.}
\end{figure}

We will now explore the threshold at which the ancilla-assisted scheme outperforms the other states involving squeezing. 
In Fig.~\ref{fig:su11_anc} we plot the minimum QFI difference between the ancilla-assisted strategy and the SU(1,1) interferometer (Fig~\ref{fig:su11} b) for a range of values of $r$ and $\eta$. That is, we plot the inverse of Eq.~\eqref{eq:tmsv} subtracting Eq.~\eqref{eq:qfi_su11}.
 Here the mean photon numbers going through the phase shift are taken to be equal. We see that the ancilla-assisted scheme is the most advantageous in the high-loss and high photon-number regime. 

\begin{figure}
\includegraphics[trim = 0cm 0cm 0cm 0cm, clip, width=0.9\linewidth]{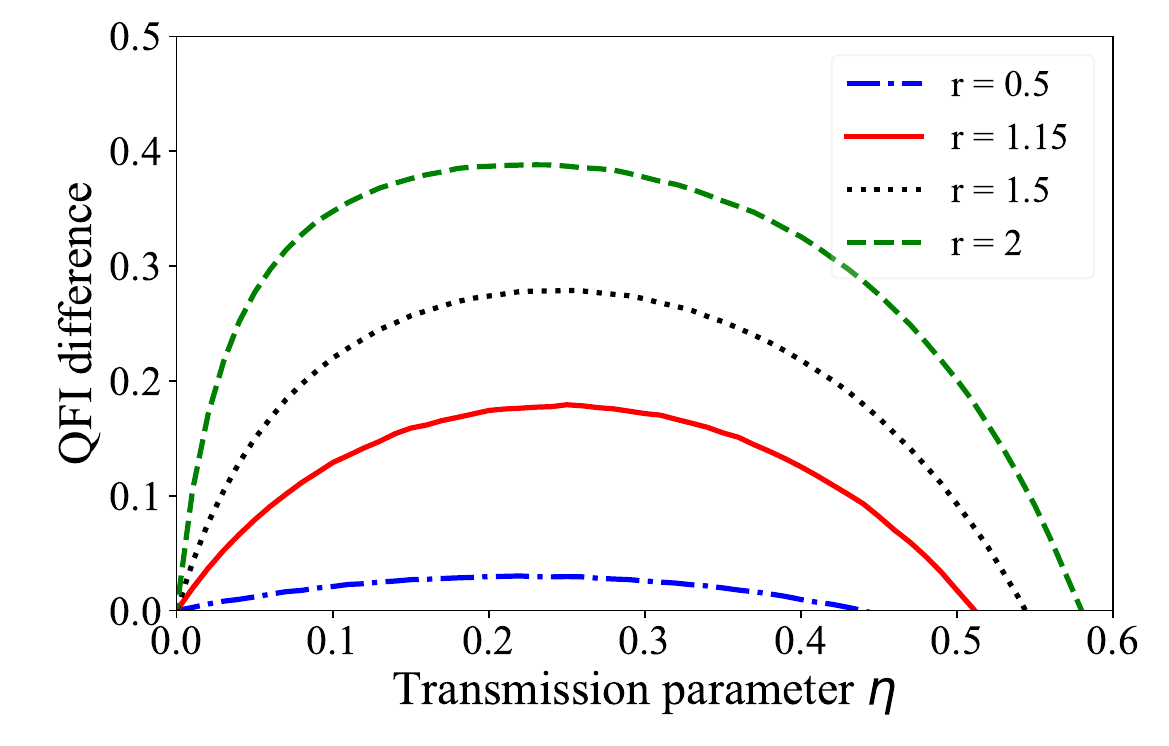}
\caption{\label{fig:caves_anc} The minimum QFI difference between the ancilla-assisted scheme and the Caves scheme in Fig~\ref{fig:caves_qfi} for a number of values of the squeezing parameter.}
\end{figure}

 We note a similar behaviour in Fig.~\ref{fig:caves_anc} where we plot the QFI difference between the ancilla-assisted strategy and the Caves scheme for a few different values of photon number. We show this in 2D for clarity because the results are numerical.

\begin{figure}
\includegraphics[trim = 0cm 0cm 0cm 0cm, clip, width=1.0\linewidth]{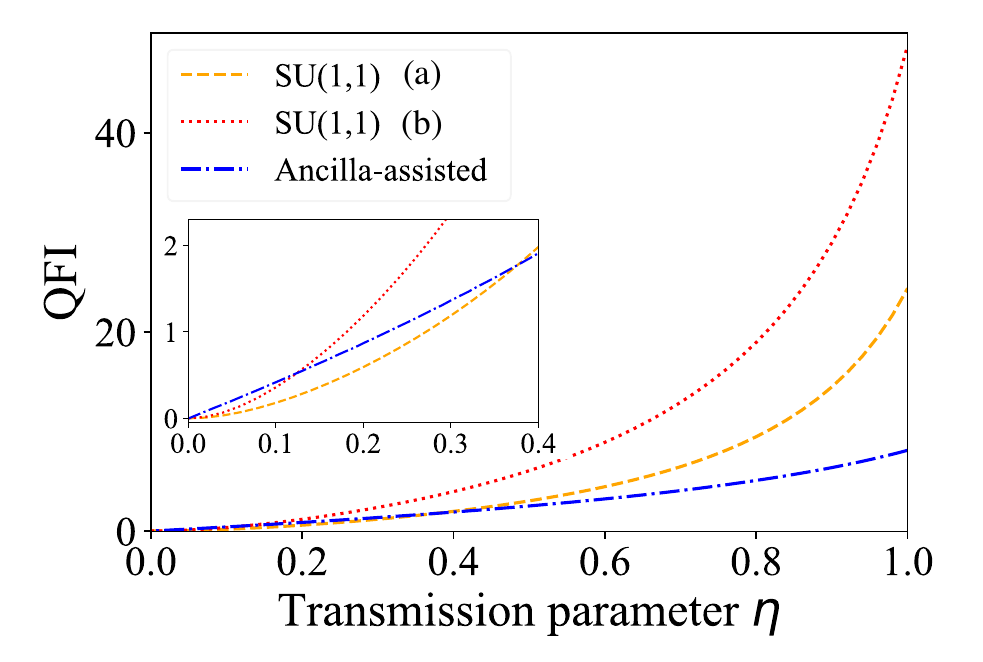}
\caption{\label{fig:squeezerfix} QFI for an interferometer with total photon number produced by squeezing $\bar N =2\sinh^2r$, $r=1.15$: the
SU(1,1) interferometer with added BS (red dotted line), the SU(1,1) interferometer (yellow dashed line) The ancilla-assisted state as in Fig.~\ref{fig:tmsv} (blue dotted-doshed line, the squeezing parameter here is also $r=1.15$).}
\end{figure}

Finally, we compare the ancilla-assisted scheme with the SU(1,1) interferometer, fixing the total number of photons coming from the squeezers instead of the mean photon number going through the phase shift (since squeezing is often considered as a much more precious resource). The 
comparison is seen in Fig.~\ref{fig:squeezerfix}. We see a similar behaviour to previous figures: in the high-loss regime, the ancilla-assisted strategy out-performs the SU(1,1) interferometers. This advantage persists despite the phase shift experiencing less number of photons passing through.

 The improvement by using an ancilla-assisted scheme is reminiscent of Gaussian quantum illumination protocols \cite{PhysRevA.71.062340,lloyd2008enhanced,PhysRevLett.101.253601}.
Unlike other quantum parameter estimation and sensing strategies where noise quickly destroys the quantum enhancement \cite{guta,PhysRevLett.113.250801}, quantum illumination retains its advantage over classical strategies, even when the entanglement has been completely destroyed \cite{PhysRevLett.101.253601,PhysRevLett.114.110506}. Analogously, here we see that our scheme retains an advantage in the presence of high loss. 
  In fact, our protocol uses the same initial state as Gaussian quantum illumination. Therefore one can explain the origin 
  of the enhancement of the ancilla-assisted protocol --- the ancilla-probe system correlation increases phase estimation precision, as evident in the observable $(b_1 b_3 - b_1^\dagger b_3^\dagger)$ 
  in Eq.~\eqref{eq:anc_obs}. 

The optimal measurement corresponds to a number measurement in the Fock basis that diagonalises the SLD \cite{serafini2017quantum}. However, even if one can diagonalise the SLD,
translating that into a physical implementation is non-trivial. 
The physical measurement will likely involve a series of beam splitters, squeezers and ancillary modes, followed by photon counting and homodyne measurements. \color{black}
While the observables presented in this paper are not easily implemented in practice in the laboratory, procedures that approximate them might be possible. This will be analysed in future work. 
It is worth noting that the minimum phase variance for a coherent state input is 
$\Delta^2\varphi \geq 1/(2 \alpha^2)$, which is a factor of 2 larger than the benchmark $\Delta^2 \varphi \geq 1/\alpha^2$. The $ 1/(2 \alpha^2)$ precision is achievable with homodyne measurement on the mode with
a strong local oscillator whose position/momentum is precisely defined. Note that this requires a local oscillator with infinite power (see Appendix~\ref{append_a}).


\section{Conclusion}

In conclusion, we have shown that an ancilla-assisted strategy for Gaussian state interferometry can beat the coherent state for all levels of the loss parameter. In the high-loss regime, we see that this strategy out-performs many schemes proposed to date. We also show that the quantum Fisher information of the Caves scheme is larger than $\alpha^2 e^{2r} + \sinh^2 r$, quoted in all current literature, and there exists a measurement that saturates the bound. By using the appropriate measurement, the precision of the Caves scheme can be improved upon, and is more loss-resistant than photon counting. For all the
schemes we examined, we have found the optimal measurement operators.


\appendix*

\section{Appendix}

\subsection{Example calculation}
\label{append_cal}

Calculating the variance of the observables involve taking the the second and fourth moments of creation and annihilation operators on the respective modes. This was done by calculating the evolution of the operators in the Heisenberg picture, then taking the vacuum expectation value.
For example, to calculate the variance of the operator $\hat Y = a_1 a_2 - a_1^\dagger a_2^\dagger$ for the TMSV, we obtain:

\begin{widetext}
\begin{align}
\braket{\hat Y}&=\bra{0,0}_{1,2} \hat S(r)^\dagger \hat Y \hat S(r) \ket{0,0}_{1,2} \nn
S(r) &= \exp(r^* a_1 a_2 - r a_1^\dagger a_2^\dagger) \nn
\braket{\hat  Y^2}&=\bra{0,0}_{1,2} \hat S(r)^\dagger 
(a_1 ^2 a_2 ^2 - a_1 a_2 a_1^\dagger a_2 ^\dagger -a_1^\dagger a_2 ^\dagger a_1 a_2  +
a_1 ^{\dagger 2} a_2 ^{\dagger 2})\hat S(r) \ket{0,0}_{1,2}
\end{align}
\noindent The non-commutative algebra was simplified using the Quantum Computing Mathematica package.

For the Caves-scheme variant, the expectation value and second moment of the observable $\hat M$ in Eq.~\eqref{eq:caves_obs}, when $\eta=1$ become

\begin{align}
\braket{\hat  M}&=\frac{1}{8} e^{-4 r} \sin (\theta -\phi ) \left(\left(e^{4 r}-1\right) \left(4 \alpha ^2 e^{2 r}+e^{4 r}-1\right) \cos (\theta -\phi )+2 e^{2 r} \left(2 \alpha ^2+\left(2 \alpha ^2+1\right) e^{4 r}+\left(4 \alpha ^2-2\right) e^{2 r}+1\right)\right)\nn
\braket{\hat M^2} &= \frac{1}{512} e^{-8 r}(80 e^{4 r} \alpha ^4+256 e^{6 r} \alpha ^4+352 e^{8 r} \alpha ^4+256 e^{10 r} \alpha ^4+80 e^{12 r} \alpha ^4+64 e^{4 r} \cos (3 (\theta -\phi )) \alpha ^4\nn
&+128 e^{6 r} \cos (3 (\theta -\phi )) \alpha ^4-128 e^{10 r} \cos (3 (\theta -\phi )) \alpha ^4-64 e^{12 r} \cos (3 (\theta -\phi )) \alpha ^4-16 e^{4 r} \cos (4 (\theta -\phi )) \alpha ^4\nn
&+32 e^{8 r} \cos (4 (\theta -\phi )) \alpha ^4-16 e^{12 r} \cos (4 (\theta -\phi )) \alpha ^4+24 e^{2 r} \alpha ^2+208 e^{4 r} \alpha ^2+280 e^{6 r} \alpha ^2+64 e^{8 r} \alpha ^2\nn
&+168 e^{10 r} \alpha ^2+240 e^{12 r} \alpha ^2+40 e^{14 r} \alpha ^2-48 e^{2 r} \cos (3 (\theta -\phi )) \alpha ^2-144 e^{6 r} \cos (3 (\theta -\phi )) \alpha ^2\nn
&+192 e^{8 r} \cos (3 (\theta -\phi )) \alpha ^2+240 e^{10 r} \cos (3 (\theta -\phi )) \alpha ^2-192 e^{12 r} \cos (3 (\theta -\phi )) \alpha ^2-48 e^{14 r} \cos (3 (\theta -\phi )) \alpha ^2\nn
&+24 e^{2 r} \cos (4 (\theta -\phi )) \alpha ^2-72 e^{6 r} \cos (4 (\theta -\phi )) \alpha ^2+72 e^{10 r} \cos (4 (\theta -\phi )) \alpha ^2-24 e^{14 r} \cos (4 (\theta -\phi )) \alpha ^2+12 e^{2 r}\nn
&+60 e^{4 r}-12 e^{6 r}-126 e^{8 r}-12 e^{10 r}+60 e^{12 r}+12 e^{14 r}+3 e^{16 r}\nn
&+8 e^{2 r} \left(-1+e^{4 r}\right) [8 e^{4 r} \left(2 \alpha ^2+1\right) \alpha ^2 +2 \alpha ^2+e^{8 r} \left(6 \alpha ^2+1\right)+e^{6 r} \left(8 \alpha ^4+24 \alpha ^2-2\right)\nn
&\hspace{3mm}+e^{2 r} \left(8 \alpha ^4+24 \alpha ^2+2\right)-1] \cos (\theta -\phi )\nn
&-4 e^{2 r} [-4 \alpha ^2+e^{12 r} \left(4 \alpha ^2-1\right)+2 e^{2 r} \left(8 \alpha ^4+2 \alpha ^2+5\right)+2 e^{10 r} \left(8 \alpha ^4+30 \alpha ^2+5\right)+e^{4 r} \left(64 \alpha ^4+4 \alpha ^2-31\right)\nn
&\hspace{3mm} +e^{8 r} \left(64 \alpha ^4-4 \alpha ^2-31\right)+e^{6 r} \left(96 \alpha ^4-64 \alpha ^2+44\right)-1] \cos (2 (\theta -\phi ))\nn
&-24 e^{2 r} \cos (3 (\theta -\phi ))+48 e^{4 r} \cos (3 (\theta -\phi ))+24 e^{6 r} \cos (3 (\theta -\phi ))-96 e^{8 r} \cos (3 (\theta -\phi ))+24 e^{10 r} \cos (3 (\theta -\phi ))\nn
&+48 e^{12 r} \cos (3 (\theta -\phi ))-24 e^{14 r} \cos (3 (\theta -\phi ))+12 e^{4 r} \cos (4 (\theta -\phi ))-18 e^{8 r} \cos (4 (\theta -\phi ))+12 e^{12 r} \cos (4 (\theta -\phi ))\nn
&-3 e^{16 r} \cos (4 (\theta -\phi ))-3 \cos (4 (\theta -\phi ))+3)
\end{align}

This leads to the relatively simple form
\begin{eqnarray}
\Delta^2 M &=& \braket{\hat  M^2}- \braket{\hat M}^2 \nn
&=&\alpha ^2 \sin (\theta -\phi )-\frac{1}{2} \sin (\theta -\phi )-\frac{1}{4} \sin (\theta -\phi ) \cos (\theta -\phi )\nn
&&+\frac{1}{2} \alpha ^2 e^{-2 r} \sin (\theta -\phi )+\frac{1}{2} \alpha ^2 e^{2 r} \sin (\theta -\phi )-\frac{1}{2} \alpha ^2 e^{-2 r} \sin (\theta -\phi ) \cos (\theta -\phi )\nn
&&+\frac{1}{2} \alpha ^2 e^{2 r} \sin (\theta -\phi ) \cos (\theta -\phi )+\frac{1}{4} e^{-2 r} \sin (\theta -\phi )+\frac{1}{4} e^{2 r} \sin (\theta -\phi )\nn
&&+\frac{1}{8} e^{-4 r} \sin (\theta -\phi ) \cos (\theta -\phi )+\frac{1}{8} e^{4 r} \sin (\theta -\phi ) \cos (\theta -\phi ).
\end{eqnarray}
\end{widetext}

\noindent When we calculate $\Delta^2\varphi = \frac{\Delta^2 M}{(\partial \braket{M}/\partial \varphi)^2}$, we arrive at Eq.~\eqref{eq:caves_qcr} in the main text.

\subsection{Achieving the QCR bound for the coherent state}
\label{append_a}

For a coherent state input in Fig.~2,the QFI reduces to $2\alpha^2$. The optimal measurement is homodyne on the mode which experiences the phase shift. This can be implemented by measuring the position $\hat x_2$ of the state. 
\begin{eqnarray}
\hat x_2  &=& \frac{1}{\sqrt2}( a_2+ a_2^\dagger)  \nn 
\hat x_2\ket{\psi_\phi} &=& \frac{1}{\sqrt2}( a_2+ a_2^\dagger)  
\ket{\frac{\alpha}{\sqrt2},\frac{\alpha e^{i\phi}}{\sqrt2}}_{1,2}\\
\braket{\hat x_2}&=&  \frac{1}{\sqrt2}\left(\frac{\alpha e^{i\phi}}{\sqrt2}+ \frac{\alpha e^{-i\phi}}{\sqrt2} \right) \nn
&=& \alpha \cos(\phi),
\end{eqnarray}
\noindent Omitting the subscript 2 for brevity.
\begin{eqnarray}
\hat x^2 &=& \frac{1}{2}(a^2 + a a^\dagger + a^\dagger a + a^\dagger a^\dagger)\ket{\frac{\alpha e^{i\varphi}}{\sqrt2}} \nn
&=& \frac{1}{2}\left(\frac{\alpha^2 e^{2 i\phi}}{2} + \frac{\alpha^2}{2} +1 + \frac{\alpha^2}{2} + \frac{\alpha^2 e^{-2 i\phi}}{2}\right)\nn
&=& \frac{1}{2}\left(\alpha^2 \cos(2\phi) + \alpha^2 +1 \right) \nn
&=&\frac{1}{2}[\alpha^2 (2\cos^2\phi-1) + \alpha^2 +1 ] \nn
&=&\alpha^2 (\cos^2\phi) +1/2.
\end{eqnarray}
We obtain
\begin{eqnarray}
\Delta^2 x &=& \braket{\hat x^2} - \braket{\hat x}^2\nn
\Delta^2\phi &=& \frac{\Delta^2 x}{(\partial x/\partial \varphi)^2} \nn
&=& \frac{1/2}{\alpha^2 \sin^2\phi} ,
\end{eqnarray}
\noindent which is optimal when $\phi=\pi/2$.

\subsection{Width of the Fisher information peak}\label{sec:width}

\begin{figure}
\includegraphics[trim = 0cm 0cm 0cm 0cm, clip, width=0.9\linewidth]{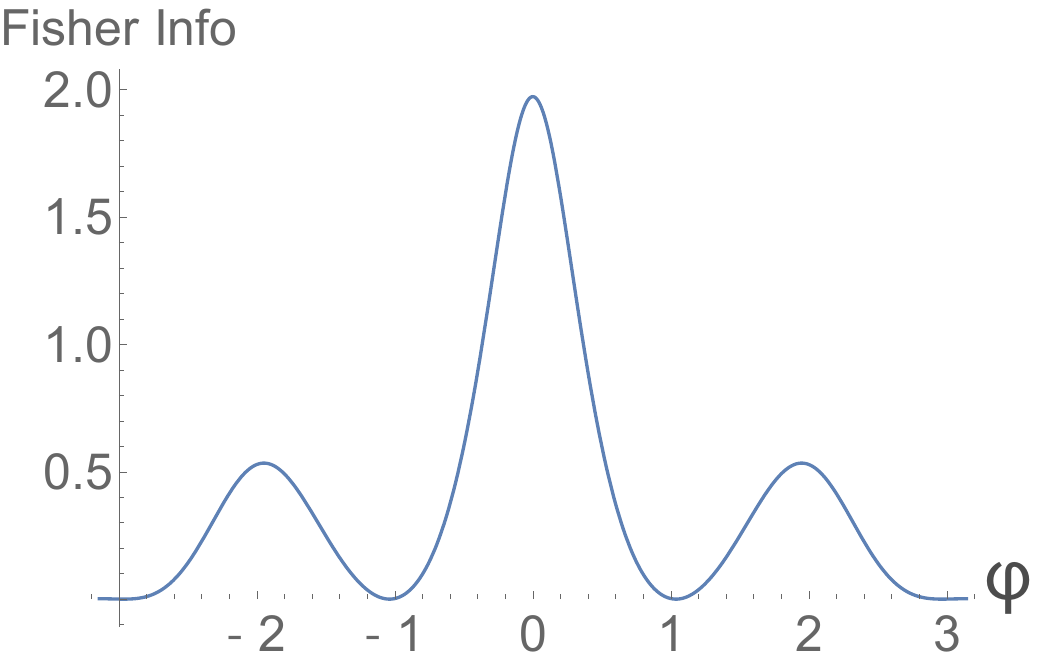}
\caption{\label{fig:qfitheta}Fisher information of the Caves-variant scheme as a 
function of the parameter $\varphi$, with parameters $r=1, \alpha=\sinh r$. }
\end{figure}

In general, the quantum Fisher information depends on the (unknown)
parameter $\varphi$ to be estimated. The Fisher information as a
function of $\varphi$ will typically have a peak where it is large
around an optimal estimation point, e.g.~in Fig.~\ref{fig:qfitheta}.
In order for the procedure to be used in an iterative manner, we need
to require that the width of the Fisher information peak is not too
narrow. This ensures that the estimation $\varphi_k$ obtained at the
$k$th iteration of the protocol can be used as a seed for the $k+1$
iteration of the protocol. Indeed, if the estimation $\varphi_k$ is
outside the peak of the Fisher information, namely where the Fisher
information is low, the successive iteration will give a very low
information and the procedure will not converge.  In contrast, if the
estimation $\varphi_k$ is in the region of high Fisher information the
$k+1$ step gives a large information on the parameter and the
successive estimation will remain in the peak region also for the
successive iteration (at least with high probability). In this case,
the procedure converges and can be iterated to large $n$.

As an example, we examine how sensitive the Caves-variant scheme
Fisher information is to the parameter: as soon as $\varphi$ moves
away from the optimal point, the Fisher information drops (see
Fig.~\ref{fig:qfitheta}). Then, if we want to iterate the procedure,
we need to ensure that we start with the interferometer in the region
of large QFI (the central peak). Namely, we need some prior
information on the phase (perhaps obtained with a classical strategy,
see e.g.~\cite{PhysRevLett.85.5098,PhysRevLett.107.233601,PhysRevA.95.053837}). The prior information at the first step of
the protocol is just a constant overhead, but to ensure that the
procedure converges when we iterate it, we must ensure that the
central peak does not shrink too quickly as a function of $n$, for the
successive iterations of the protocol.

We consider the full-width half maximum (FWHM) $W$ 
of the Fisher information peak and require that the error on the
estimation of each step of the protocol satisfies, 
\begin{align}
  \Delta \varphi &\leqslant W
\end{align}
\noindent in order for the estimation to return a high information on
$\varphi$.  Now, because the QFI of the scheme scales as $n^2$, we
need the FWHM (the bound on $\Delta\varphi$) to contract no faster
than $1/n$,
\begin{align}
  W \times n \geq \text{constant} .
  \label{eq:Wn}
\end{align}

\begin{figure}
\includegraphics[trim = 0cm 0cm 0cm 0cm, clip, width=0.9\linewidth]{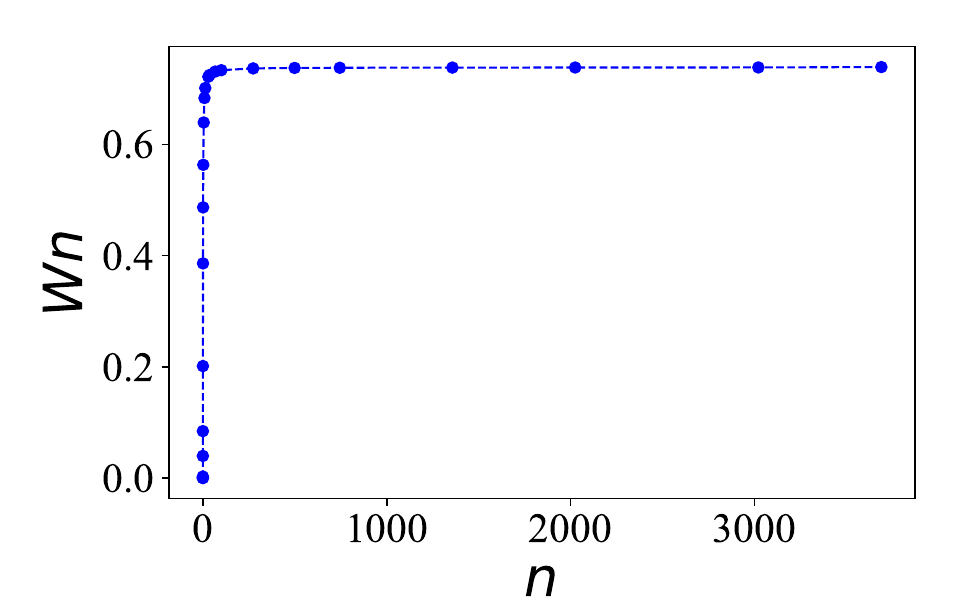}
\caption{\label{fig:w_scale} Width of the full-width half-maximum of the Caves variant
scheme scaled by mean photon number $n$. The line is a guide for the eye only.}
\end{figure}

If Eq.~\eqref{eq:Wn} holds, then at any iteration
step, the uncertainty of this step is smaller than the peak of the
maximum of the Fisher information, and the procedure will converge.
In Fig.~\ref{fig:w_scale} we plot $W\times n$ against $n$. We see that asymptotically
the condition in Eq.~\eqref{eq:Wn} holds, and therefore with a suitable adaptive strategy
the scheme can achieve Heisenberg scaling.

\begin{acknowledgements}
This research was funded in part by EPSRC Quantum Communications Hub, Grant No. EP/M013472/1. We acknowledge funding
from the University of Pavia ``Blue sky'' project - grant
n. BSR1718573.
\end{acknowledgements}

%






\end{document}